\documentclass[pdflatex,sn-aps,iicol]{sn-jnl}

\usepackage{graphicx}%
\usepackage{multirow}%
\usepackage{amsmath,amssymb,amsfonts}%
\usepackage{amsthm}%
\usepackage{mathrsfs}%
\usepackage[title]{appendix}%
\usepackage{xcolor}%
\usepackage{textcomp}%
\usepackage{manyfoot}%
\usepackage{booktabs}%
\usepackage{algorithm}%
\usepackage{algorithmicx}%
\usepackage{algpseudocode}%
\usepackage{listings}%
\usepackage{siunitx}

\begin{document}

\title[]{Optimizing Aperture Geometry in THz-TDS for Accurate Spectroscopy of Quantum Materials}

\author[]{\fnm{Laura O.} \sur{Dias}}
\equalcont{These authors contributed equally to this work.}

\author[]{\fnm{Eduardo D.} \sur{Stefanato}}
\equalcont{These authors contributed equally to this work.}

\author[]{\fnm{Nicolas M.} \sur{Kawahala}}

\author*[]{\fnm{Felix G. G.} \sur{Hernandez}}\email{felixggh@if.usp.br}

\affil[]{\orgdiv{Instituto de Física}, \orgname{Universidade de São Paulo}, \orgaddress{\city{São Paulo}, \postcode{05508-090}, \state{SP}, \country{Brazil}}}

\abstract{Terahertz time-domain spectroscopy (THz-TDS) provides a powerful platform for investigating low-energy excitations in quantum materials. Because these materials are often limited in size, experimental setups typically rely on tightly focused beams and metallic holders with small apertures. In this work, we perform a systematic study of how aperture geometry influences THz signal transmission in a standard free-space configuration. By analyzing time- and frequency-domain data for circular apertures of varying diameters and thicknesses, we quantify the spatial and spectral filtering effects imposed by aperture size. We show that small apertures progressively attenuate low-frequency components of the transmitted signal, while higher-frequency content remains comparatively unaffected. These effects become especially significant for apertures smaller than typical THz beam waists, resulting in amplitude suppression and phase distortions that compromise the accuracy of frequency-domain analysis and optical parameter retrieval. To validate these observations, additional measurements were performed on a representative quantum material, confirming the practical relevance of the identified aperture effects. The transmitted intensity as a function of aperture diameter also provides a straightforward method for estimating the beam waist at the focus. In contrast, standard aperture thicknesses do not introduce measurable distortions, confirming the adequacy of treating thick, non-resonant apertures as dielectric slabs. These findings establish practical guidelines for aperture selection in THz-TDS and underscore the importance of preserving low-frequency response for reliable characterization of quantum materials.}


\maketitle

\section{Introduction}\label{sec:intro}
Recent advances in coherent terahertz (THz) emission and detection technologies have led to the widespread adoption of terahertz time-domain spectroscopy (THz-TDS) as a powerful tool for investigating quantum materials and other condensed matter systems~\cite{Leitenstorfer2023,Lee2009}. Owing to its broadband nature---often spanning substantial portions of the \qtyrange{.1}{10}{\THz} range---THz-TDS enables direct access to low-energy excitations such as phonons~\cite{Handa2024,Hernandez2023,Kawahala2023paper,Baydin2022}, magnons~\cite{Kurihara2023,Huang2024,Zhang2024magnon,Zhang2024modes}, and excitons~\cite{Kaindl2009,Zhao2020,Haque2024}. These capabilities are especially valuable for exploring emergent properties in systems with strong correlations, topological order, or reduced dimensionality~\cite{Bera2021old}. THz-TDS also plays an important role in applications spanning electronics and photonics~\cite{Chen2019,Rogalin2018}, and is well suited for experiments under extreme physical conditions, further motivating continued improvements in its implementation.

In transmission-geometry THz-TDS, samples are typically mounted onto metallic holders featuring circular apertures, which define the illuminated region and facilitate alignment. Although this geometry is common when probing a range of material systems---from bulk~\cite{Kawahala2025,Jeon2012} to thin films~\cite{Kawahala2023paper,stefanato2025} and 2D materials~\cite{Huang2025,Wu2024}---the aperture diameter must be carefully selected to avoid undesirable optical effects. When the aperture is significantly smaller than the THz beam spot, substantial transmission losses and spectral filtering can occur. Conversely, if the aperture is too large, stray or multiply scattered rays may reach the detector, degrading signal quality and introducing spectral artifacts. Moreover, if the aperture diameter approaches the radiation wavelength, diffraction effects become prominent~\cite{Li2013,Neumann2015}. These geometric factors are therefore not merely peripheral concerns but play a central role in determining the fidelity and interpretability of THz-TDS experiments---especially in the analysis of quantum materials whose spectroscopic signatures may lie close to the detection threshold.

Small apertures at the apex of tapered metallic tips have long served as probes for near-field terahertz imaging, enabling sub-wavelength spatial resolution \cite{Hunsche1998_IEICE,Hunsche1998_OptComm}. Conversely, regarding far-field experiments, Kumar \textit{et al.}~\cite{Kumar2025} demonstrated that small apertures, such as pinholes, can act as frequency filters, reducing the detectable bandwidth and introducing distortions in THz imaging. In contrast to conventional monochromatic optics, where the beam waist is fixed by the wavelength, the broadband nature of THz pulses means that each frequency component experiences a different degree of focusing. Consequently, the beam spot size at the aperture becomes frequency-dependent, making the transmission spectrum sensitive to the aperture geometry.

The frequency-dependent beam spot at the aperture has direct implications for quantitative THz spectroscopy, especially in studies targeting the intricate low-energy physics of emerging quantum materials. These systems often require accurate retrieval of optical conductivity or dielectric response over a broad frequency range, placing stringent demands on spectral fidelity. For instance, in thin-film samples exhibiting free-carrier dynamics~\cite{Ulbricht2011,Shi2024,Song2025,Chen2025}, the low-frequency components of the pulse—most sensitive to conductivity—are particularly vulnerable to diffraction and attenuation when the wavelength approaches the aperture diameter. Such spectral distortions can compromise the extraction of material parameters and lead to misinterpretations. Depending on the spectral range of interest, a lower bound on sample dimensions may therefore be necessary to ensure measurement reliability. Understanding the impact of aperture geometry is thus critical when designing THz spectroscopy experiments.

In this study, we perform a systematic analysis of how aperture geometry influences terahertz transmission in a standard THz-TDS setup. Circular metallic apertures with varying diameters and thicknesses are positioned at the beam focus, and the transmitted pulses are recorded under identical alignment conditions. By combining time-domain and spectral analyses, we extract the beam waist at the focal plane and assess how the aperture size affects the transmitted signal across the frequency range. While direct beam-imaging techniques, such as bolometric cameras~\cite{Oda2010,Oda2015}, offer comprehensive spatial characterization, our approach provides a practical and accessible alternative based on the beam-blocking principle of knife-edge techniques~\cite{Suzaki1975,Firester1977,Cardel2013} for estimating beam parameters and identifying spectral distortions. The results offer direct guidance for optimizing aperture selection and sample mounting in high-fidelity THz-TDS measurements---particularly in studies of quantum materials, where preserving low-frequency information is essential. To illustrate the practical implications of these findings, the analysis is further extended to measurements on a representative quantum material using sample holders with different aperture diameters.

\section{Methods}\label{sec:methods}
This section describes the experimental and analytical procedures employed to investigate the terahertz response of circular apertures in a standard THz-TDS setup. We begin by outlining the fabrication and geometrical characterization of the apertures, followed by details of the time-domain spectroscopy measurements. Finally, we present the theoretical model used to analyze the transmission spectra and extract optical parameters.

\subsection{Apertures}\label{ssec:apertures}
The sample holders used in this study consist of metallic plates with circular apertures, a geometry well-suited for spatially symmetric Gaussian beams. In typical THz-TDS experiments, samples are mounted onto such holders so that the beam interacts with the material either before or after passing through the aperture, depending on the configuration. Here, however, the focus is on investigating the transmission through the apertures themselves in the absence of any sample, isolating the effects of aperture geometry on the detected signal. The circular shape ensures rotational symmetry and uniform beam truncation, minimizing edge diffraction and justifying its widespread use in THz spectroscopy.

We employed a set of \num{10} apertures with diameters ranging from \qtyrange{0.5}{5.0}{\mm} in uniform steps, all fabricated from \qty{2}{\mm}-thick aluminum plates. To investigate possible cavity or waveguide-like effects, an additional set of \num{4} apertures with fixed diameter (\qty{3.0}{\mm}) and varying thicknesses between \qty{1}{\mm} and \qty{6}{\mm} was also studied.

\subsection{THz-TDS}\label{ssec:thz-tds}
\begin{figure*}[t]
\centering
\includegraphics[width=0.94\linewidth]{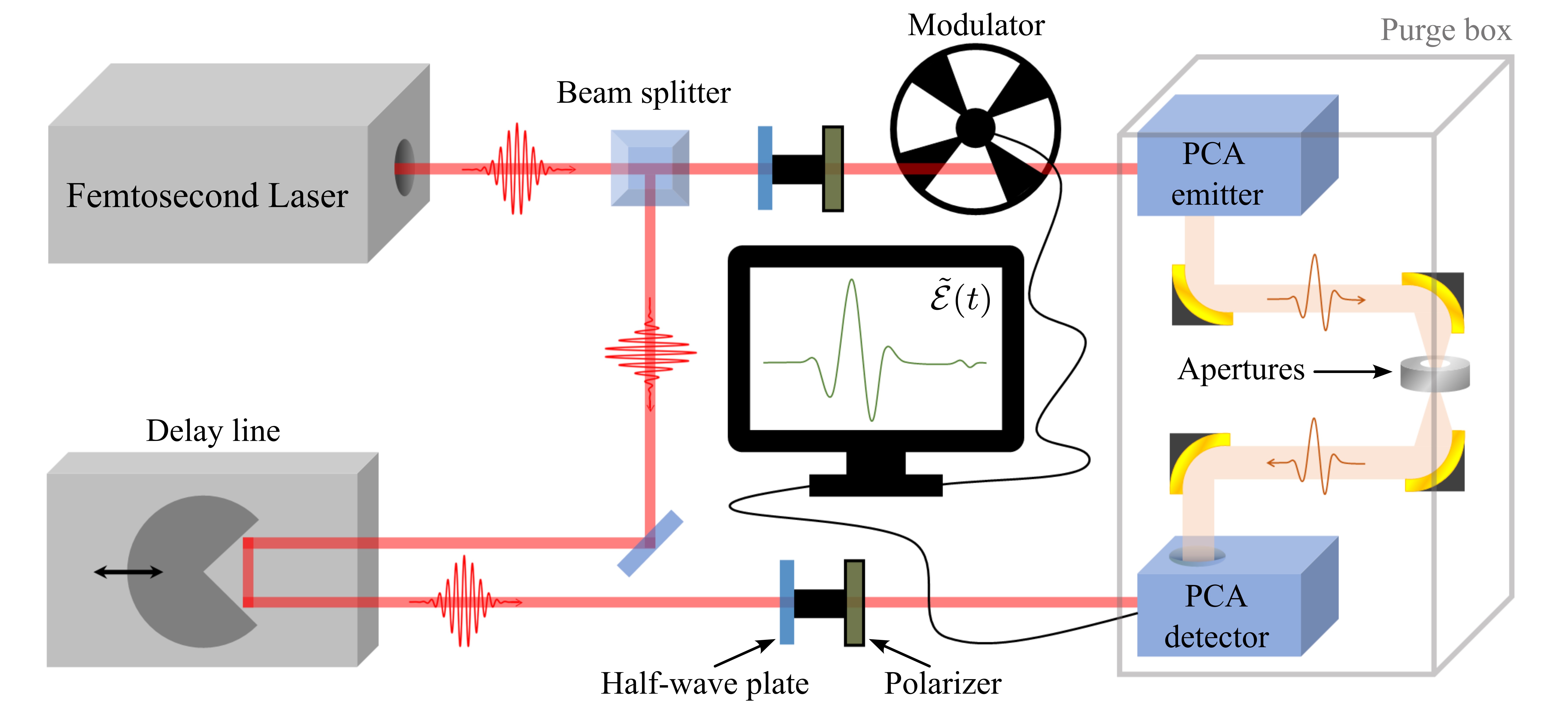}
\caption{Schematic diagram of the THz-TDS setup. Infrared laser pulses are split into two optical paths: one directed to a PCA emitter, which generates broadband THz radiation, and the other routed through a mechanical delay line for temporal scanning. The THz pulse passes through the sample region---where aperture holders are positioned---and is detected by a second PCA. The transmitted electric field is recorded as a function of time.}\label{fig1}
\end{figure*}

All measurements were performed at ambient temperature and under controlled humidity conditions below \qty{10}{\percent}, typical for laboratory THz-TDS experiments. The setup operates in transmission geometry, as illustrated in Fig.~\ref{fig1}. Infrared laser pulses are split into two optical paths: one directed to a photoconductive antenna (PCA) emitter, which generates broadband THz radiation, and the other routed through a mechanical delay line that controls the temporal overlap between the THz and gating pulses.

The THz beam is focused and subsequently recollimated using a set of off-axis parabolic mirrors. The circular apertures are positioned at the beam waist (the focal point between the mirrors). Then, the beam is directed onto a second PCA for coherent detection. The time-resolved electric field is acquired by scanning the optical delay line~\cite{Lee2009}.

The time-domain signal is Fourier-transformed (e.g., via the Fast Fourier Transform~\cite{nussbaumer1982fast}) to obtain the electric field in the frequency domain, enabling spectral analysis. Because THz-TDS directly measures the electric field, both the amplitude and phase components of the transmitted signal are retrieved simultaneously. This allows for the extraction of the complex refractive index using well-established frequency-domain models~\cite{Duvillaret1996}. To improve spectral resolution, zero-padding is applied prior to transformation~\cite{Donnelle2005,marulanda2025}. In the aperture-only measurements that constitute the core of this study, no physical sample is present and air serves as reference. A brief validation with a quantum material is also included; those data are processed with the same protocol.

\subsection{Transmission model}\label{ssec:model}
To extract optical parameters from the measured signals, we adopt a transmission model commonly used in THz spectroscopy~\cite{Neu2018}. In this framework, each aperture is modeled as a cylindrical slab of homogeneous material---in this case air---with thickness $d$, while the absence of metallic boundaries serves as the reference condition. This approach enables an optical analysis of how aperture geometry influences the transmitted signal in THz-TDS experiments.

A terahertz pulse normally incident on a planar interface undergoes partial transmission and reflection, governed by the Fresnel coefficient~\cite{fresnel1870oeuvres,born2013principles}:
\begin{equation} \label{eq:2-1}
\tilde{t}_{jk} = \frac{2\tilde{n}_j}{\tilde{n}_j + \tilde{n}_k} \,,
\end{equation}
where $\tilde{n}_j$ and $\tilde{n}_k$ are the complex refractive indices of the incident and transmitted media, respectively. In this study, the aperture is treated as a dielectric slab with complex refractive index $\tilde{n}_\textrm{a}$, surrounded by air ($\tilde{n}_0=1$). The wave encounters two interfaces: from air into the aperture region (Fresnel coefficient $\tilde{t}_{0\textrm{a}}$) and back into air ($\tilde{t}_{\textrm{a}0}$).

In general, multiple internal reflections within the slab may lead to Fabry--Pérot interference~\cite{Duvillaret1996}. However, because the apertures are modeled as uniform, non-resonant air-filled volumes, we neglect such effects. We therefore consider only the propagation factor:
\begin{equation}\label{eq:propagation_factor}
    p_{d,j} = e^{2\pi i\tilde{n}_j\nu d/c},
\end{equation}
which describes both phase accumulation and amplitude attenuation over the slab thickness $d$, for a complex refractive index $\tilde{n}_j$ at frequency $\nu$. Here, $c$ is the speed of light in vacuum and $\omega=2\pi\nu$ is the angular frequency.

Under these assumptions, the frequency-dependent transmission coefficient becomes
\begin{equation} \label{eq:2-2}
    \tilde{\mathcal{T}} = \frac{\tilde{\mathcal{E}}_{\textrm{a}}}{\tilde{\mathcal{E}}_{\textrm{r}}} = \frac{\tilde{t}_{0\textrm{a}}p_{d,\textrm{a}}\tilde{t}_{\textrm{a}0}}{p_{d,0}},
\end{equation}
where $\tilde{\mathcal{E}}_\textrm{a}$ and $\tilde{\mathcal{E}}_\textrm{r}$ are the electric fields transmitted with and without the aperture, respectively.

Applying the model with $\tilde{n}_\textrm{a} = n + i\kappa$---where $n$ is the real refractive index and $\kappa$ is the extinction coefficient---and simplifying the product of Fresnel coefficients under the weak absorption approximation ($n\gg\kappa$)~\cite{jepsen_terahertz_2011}, the transmission becomes
\begin{equation}\label{eq:transmission_model}
    \tilde{\mathcal{T}}(\nu) = \frac{4n}{(n+1)^2} e^{-\kappa \nu d / c} e^{i(n-1)\nu d / c} \,,
\end{equation}
where the exponential terms account for amplitude attenuation and phase delay relative to free-space propagation.

The experimental transmission coefficient $\tilde{\mathcal{T}}_\textrm{exp}(\nu)$, computed directly from the measured fields, can be compared to Eq.~\eqref{eq:transmission_model} to retrieve the optical parameters of the aperture region. By analytically inverting the model, we obtain:
\begin{equation} \label{eq:2-8}
    \begin{cases}
    n(\nu) = 1 + \frac{c}{\nu d} \arg\left[\tilde{\mathcal{T}}_{\exp}(\nu)\right] \,, \\
    \kappa(\nu) = -\frac{c}{\nu d} \ln \left[ \frac{(n(\nu) + 1)^2}{4n(\nu)} \left|\tilde{\mathcal{T}}_{\exp}(\nu)\right| \right] \,,
    \end{cases}
\end{equation}
where $\arg[\cdot]$ and $|\cdot|$ denote the phase and amplitude of the complex transmission, respectively. These expressions, although derived here for the apertures, mirror those commonly employed in the analysis of bulk samples using THz-TDS. When combined with standard strategies such as time-domain truncation or windowing~\cite{marulanda2025}, they offer a reliable means of retrieving the complex refractive index from transmission data in a variety of experimental contexts~\cite{Koch2023}.

\section{Results and discussion}\label{sec:results}
\begin{figure*}[t]
\centering
\includegraphics[width=0.98\linewidth]{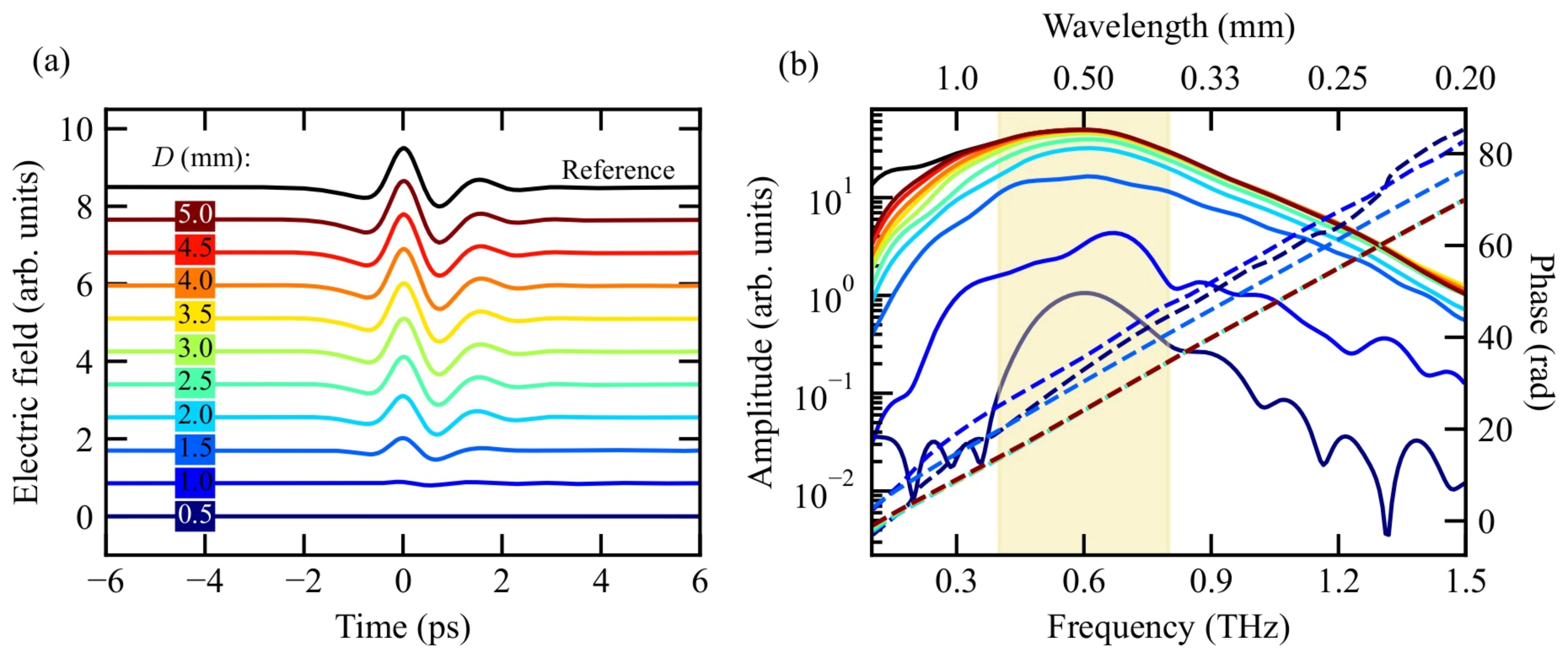}
\caption{\textbf{(a)} Time-domain electric field pulses acquired via THz-TDS for different circular aperture diameters $D$, along with a reference measurement without an aperture (black). A systematic reduction in pulse amplitude is observed with decreasing aperture size. \textbf{(b)} Corresponding amplitude (solid) and phase (dashed) spectra obtained via Fourier transform. Smaller apertures cause substantial attenuation of low-frequency components and induce phase errors. The shaded region highlights the spectral range from \qtyrange{0.4}{0.8}{\THz}.}\label{fig2}
\end{figure*}

The following results correspond to THz signals transmitted through circular apertures of varying diameters $D$. Figure~\ref{fig2}a presents the time-domain electric field profiles, vertically offset for clarity, alongside a reference pulse acquired without any aperture (black line). As the aperture diameter decreases, the transmitted signal amplitude is significantly reduced across the entire waveform, although this attenuation is not strictly linear. The corresponding amplitude and phase spectra in Fig.~\ref{fig2}b, obtained via Fourier transform, further reveal the frequency-dependent nature of this attenuation.

These measurements highlight two main trends. First, the reduction in signal amplitude reflects the spatial filtering imposed by the apertures, which increasingly block the beam as their diameter shrinks. Second, the spectral content is not uniformly suppressed: low-frequency components are more strongly attenuated, while high-frequency components are comparatively preserved---particularly for intermediate aperture sizes. This behavior reflects the frequency-dependent beam waist of the focused THz beam, where longer wavelengths result in larger spot sizes that are more strongly blocked by the apertures. Additionally, for smaller apertures, the spectral phase remains approximately linear but exhibits a noticeable vertical offset compared to larger apertures, likely due to increased phase uncertainty at low frequencies. While this shift does not affect the overall linearity, it may require correction to ensure consistent phase referencing across measurements.

We analyze these data from complementary perspectives. In the time domain, we use the relative amplitude of the transmitted pulses to extract the beam waist at the focus, providing a direct measure of spatial confinement. In the frequency domain, we examine the detailed spectral response of each aperture to characterize their frequency-dependent filtering behavior. We then apply a transmission model to extract optical parameters from the data, an important step for assessing potential challenges when retrieving material properties using small apertures in THz-TDS measurements. Additionally, we present results from an independent dataset obtained with varying aperture thicknesses to investigate possible cavity effects associated with aperture geometry. To further demonstrate the relevance of these findings, we conclude by examining THz-TDS measurements of a representative quantum material obtained using apertures of different diameters.

\subsection{Beam waist estimation}
To quantify the beam waist at the sample position, we model the THz beam intensity profile as Gaussian---consistent with the fundamental Transverse Electro-Magnetic mode (TEM$_{00}$)---and analyze the variation in transmitted intensity as a function of aperture diameter. A radial Gaussian distribution is described by~\cite{siegman1986lasers,saleh2019fundamentals}:
\begin{equation}
    I(r) = I_0e^{-2r^2/w_0^2}\,,
\end{equation}
where $w_0$ is the beam waist radius. When a circular aperture of diameter $D$ is positioned at the beam focus, the total transmitted power corresponds to the integral of this distribution over the aperture area. This yields the analytical expression~\cite{pedrotti2017introduction}
\begin{equation}\label{eq:power}
    P(D) = P_0\left(1-e^{-D^2/2w_0^2}\right)\,,
\end{equation}
where $P_0$ is the total transmitted power in the absence of clipping.

\begin{figure}[h]
\centering
\includegraphics[width=1.0\linewidth]{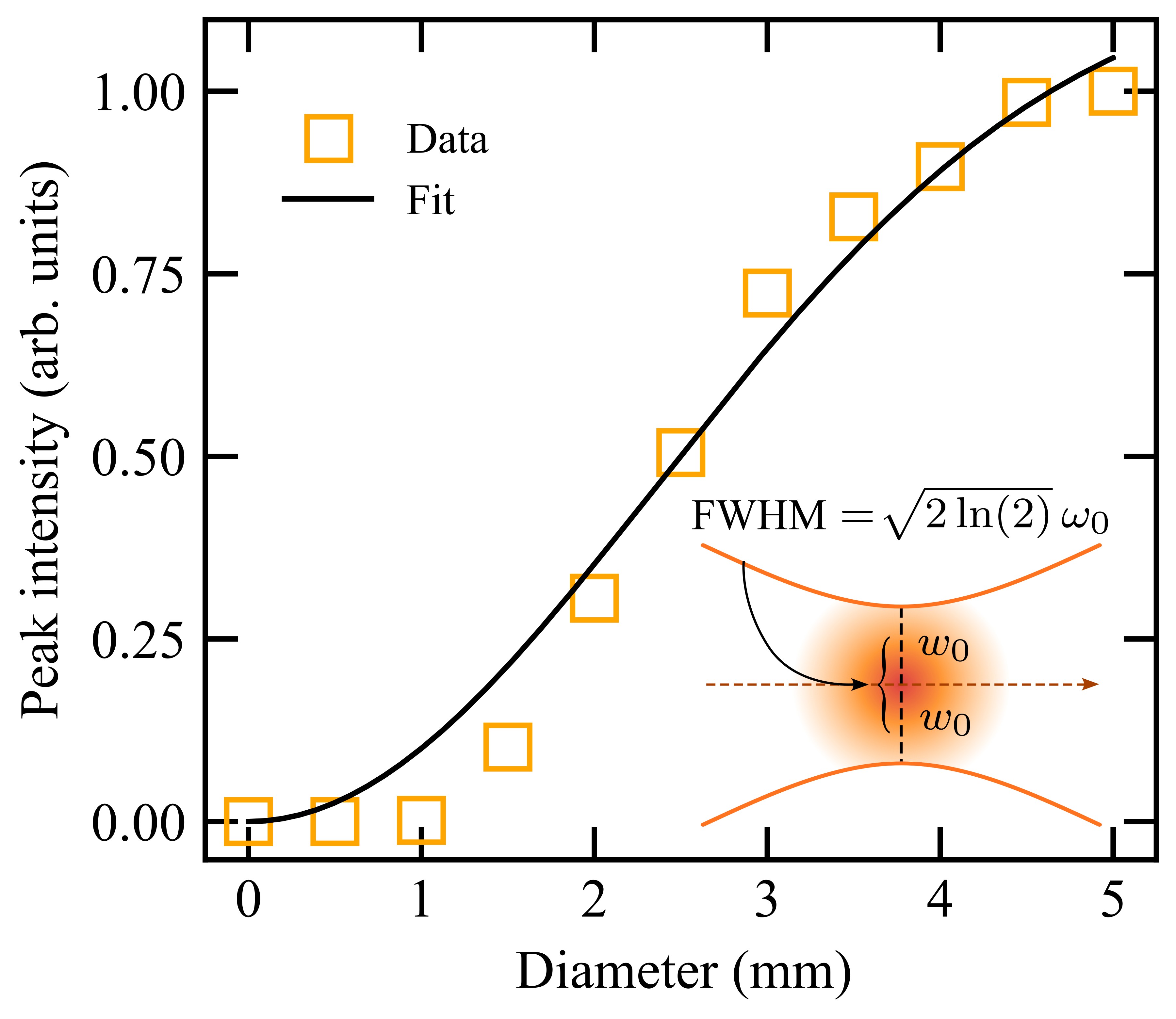}
\caption{Transmitted intensity as a function of aperture diameter. The experimental points (orange squares) represent the squared peak electric field amplitudes obtained from time-domain THz measurements, serving as a proxy for relative transmitted power. The solid black line is a fit based on the analytical model for Gaussian beam power transmission through a circular aperture [Eq.~\eqref{eq:power}]. The inset illustrates the Gaussian beam profile near the focal plane, with the beam waist $w_0$ indicated, along with its relation to the full width at half maximum (FWHM). The fit yields $w_0 = \qty{2.36(21)}{\mm}$ and a corresponding FWHM of \qty{2.78(25)}{\mm}.}\label{fig3}
\end{figure}

Fig.~\ref{fig3} shows the squared peak electric field amplitudes, extracted from the time-domain signals in Fig.~\ref{fig2}a, plotted as a function of aperture diameter. As expected, the transmitted intensity rises steeply around the beam waist and gradually saturates at larger diameters, where most of the beam propagates through the aperture without obstruction.

While Eq.~\eqref{eq:power} models the total transmitted power, the experimental observable in our case is the square of the peak electric field amplitude. This quantity is proportional to the instantaneous intensity of the pulse and, assuming the temporal pulse shape remains approximately unchanged across different apertures, provides a reliable proxy for relative transmitted power. This assumption is supported by the consistent waveform morphology observed in the time-domain signals and is commonly adopted in THz beam characterization~\cite{Reiten2003, Molloy2013}.

We fit the model in Eq.~\eqref{eq:power} to the experimental data using a nonlinear least-squares procedure. The retrieved beam waist was $w_0 = \qty{2.36(21)}{\mm}$, corresponding to a full width at half maximum (FWHM) of \qty{2.78(25)}{\mm}. The agreement between the model and the data supports the assumption of a Gaussian beam profile at the focal plane. Minor deviations at small aperture diameters may result from edge effects, emitter asymmetry, or reduced signal-to-noise ratio in the most attenuated signals~\cite{Molloy2013}.

\subsection{Spectral response of the apertures}

\begin{figure}[h]
\centering
\includegraphics[width=1.0\linewidth]{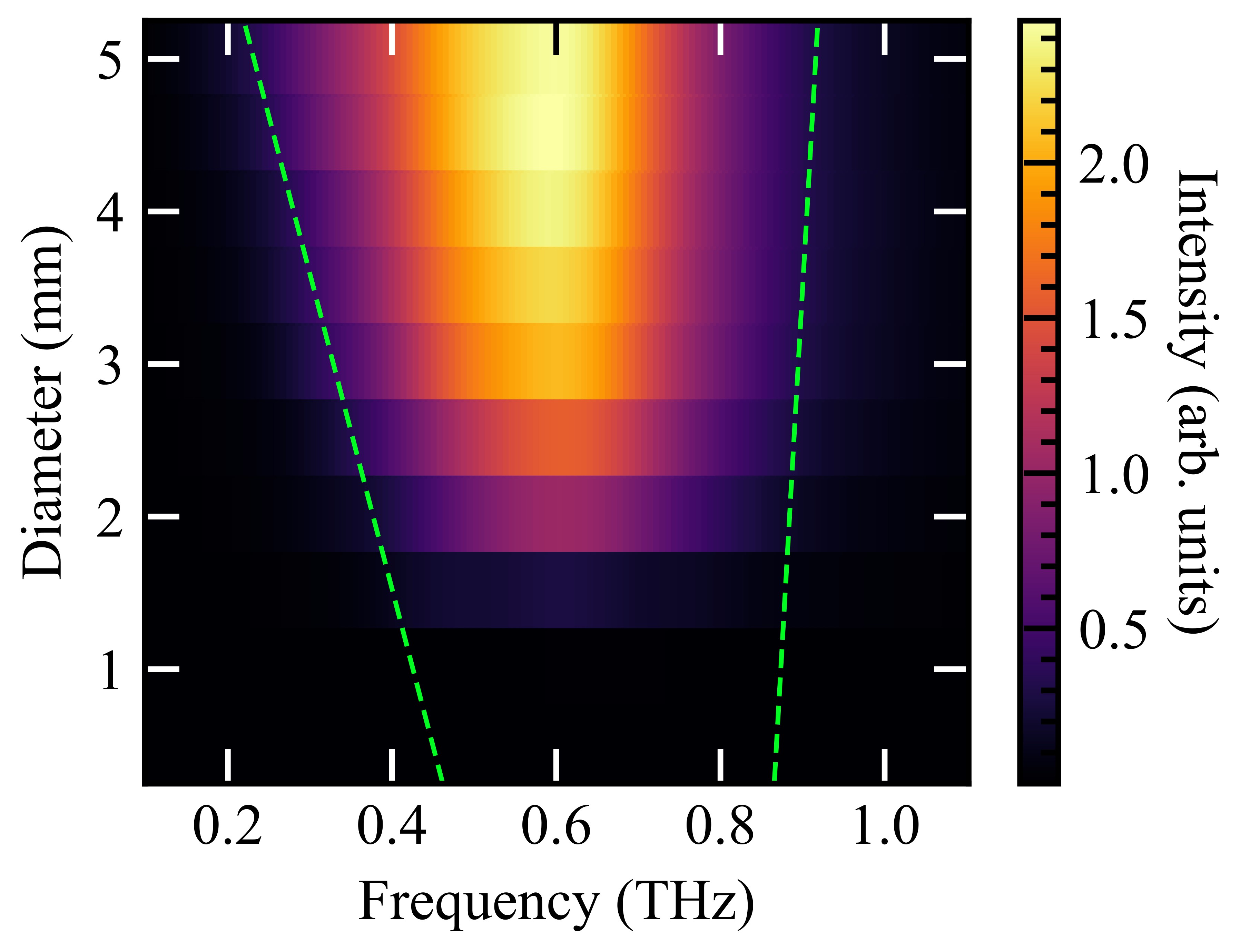}
\caption{Transmitted spectral intensity as a function of frequency and aperture diameter. The colormap is derived from Fourier transforms of the time-domain signals [Fig.~\ref{fig2}]. As the aperture diameter decreases, low-frequency components are progressively attenuated, while high-frequency components remain comparatively unaffected. Two dashed green lines are overlaid as visual guides: the left-hand curve traces the shifting low-frequency boundary of the transmitted spectrum, and the right-hand curve marks the nearly constant high-frequency edge. These features reflect the frequency-dependent spatial confinement of the focused THz beam and the corresponding filtering effect imposed by the apertures.}\label{fig4}
\end{figure}

Following the beam waist estimation from time-domain analysis, we now examine in more detail the frequency-dependent transmission characteristics of the apertures. Fig.\ref{fig4} presents a two-dimensional colormap of the transmitted spectral intensity as a function of frequency and aperture diameter, derived from the same Fourier-transformed data shown in Fig.\ref{fig2}b.

The map reveals a continuous evolution of the spectral content: as the aperture diameter decreases, the low-frequency portion of the spectrum is progressively truncated, while higher-frequency components remain relatively unaffected until very small diameters. This effect reflects the frequency-selective spatial filtering imposed by the apertures, governed by the beam’s frequency-dependent waist at the focal plane.

To aid visualization, two dashed green lines are overlaid on the map. The first traces the shifting low-frequency boundary of the transmitted spectrum, moving toward higher frequencies with decreasing aperture size. The second marks the nearly constant high-frequency edge, indicating that higher-frequency components maintain consistent transmission across aperture sizes. These guide lines reinforce the importance of aperture diameter in preserving low-frequency spectral information critical for THz-TDS applications.

\subsection{Extraction of effective optical parameters}
We apply the transmission model from the Methods section to retrieve effective optical parameters from the measured spectra. Although the apertures contain only air, each is treated as a dielectric slab of finite thickness, with transmission evaluated relative to the unobstructed reference. This approach reveals how aperture geometry alone can distort the retrieval of optical constants in THz-TDS experiments.

\begin{figure}[h]
\centering
\includegraphics[width=1.0\linewidth]{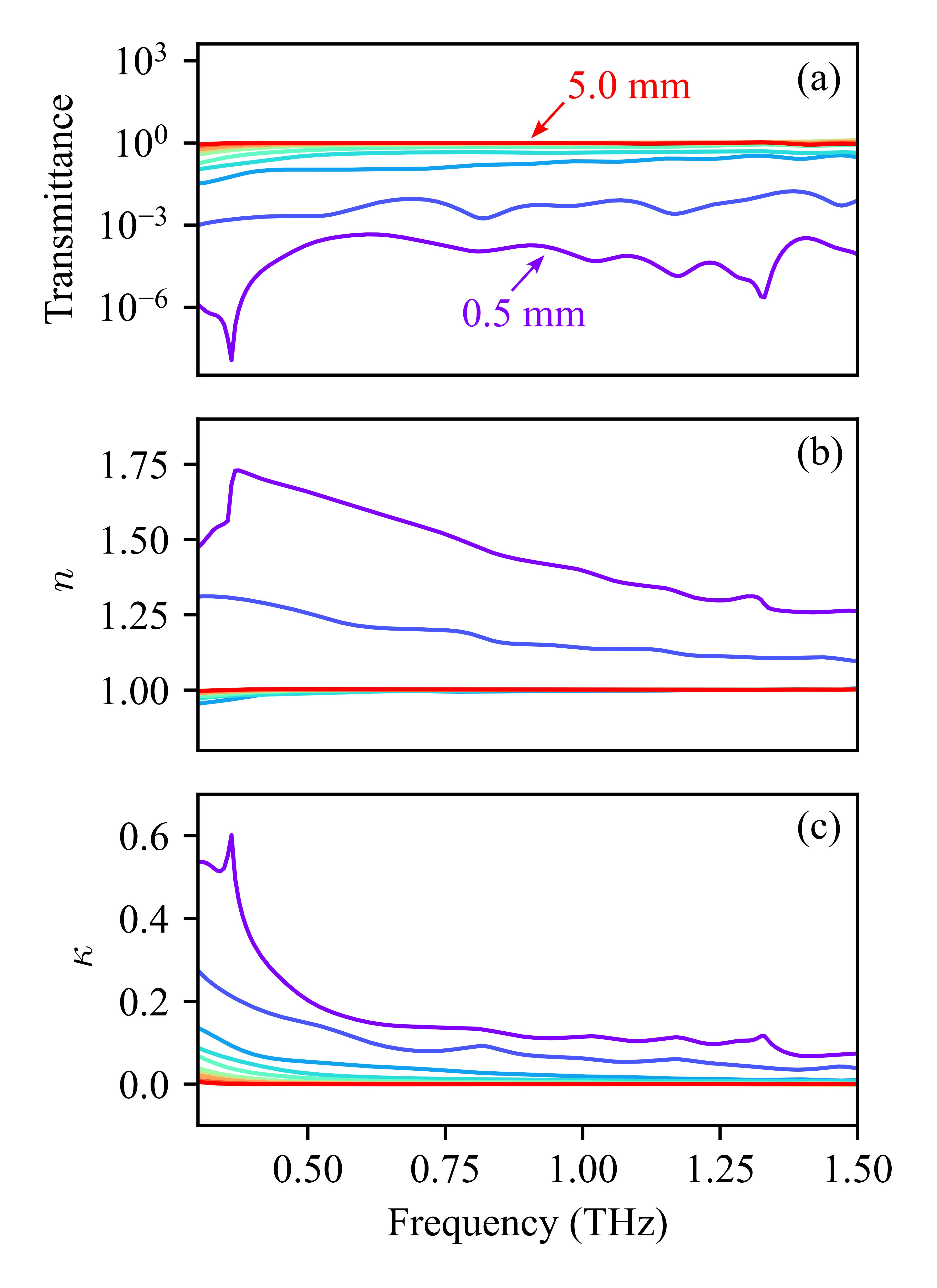}
\caption{Extracted optical parameters as a function of frequency for different aperture diameters. \textbf{(a)} Transmittance, defined as the squared modulus of the complex transmission coefficient. \textbf{(b)} Real refractive index and \textbf{(c)} extinction coefficient, both retrieved by analytically inverting the transmission model described in the Methods section. For large apertures, the extracted values converge toward the expected free-space limits ($n = 1$, $\kappa = 0$). As the diameter decreases, both parameters deviate significantly, particularly at low frequencies, reflecting the influence of beam truncation and spectral distortion on parameter retrieval in THz-TDS.}\label{fig5}
\end{figure}

Figure~\ref{fig5} presents the extracted parameters as a function of frequency: (a) transmittance, defined as  $|\tilde{\mathcal{T}}(\nu)|^2$, the squared modulus of the complex transmission coefficient; (b) real refractive index $n$; and (c) extinction coefficient $\kappa$. For larger apertures, both $n$ and $\kappa$ approach their expected free-space values ($n = 1$, $\kappa = 0$), indicating minimal perturbation to the THz beam. In contrast, smaller diameters lead to pronounced deviations—particularly at lower frequencies—reflecting the impact of beam truncation on the retrieved optical response.

\begin{figure*}[ht!]
\centering
\includegraphics[width=0.98\linewidth]{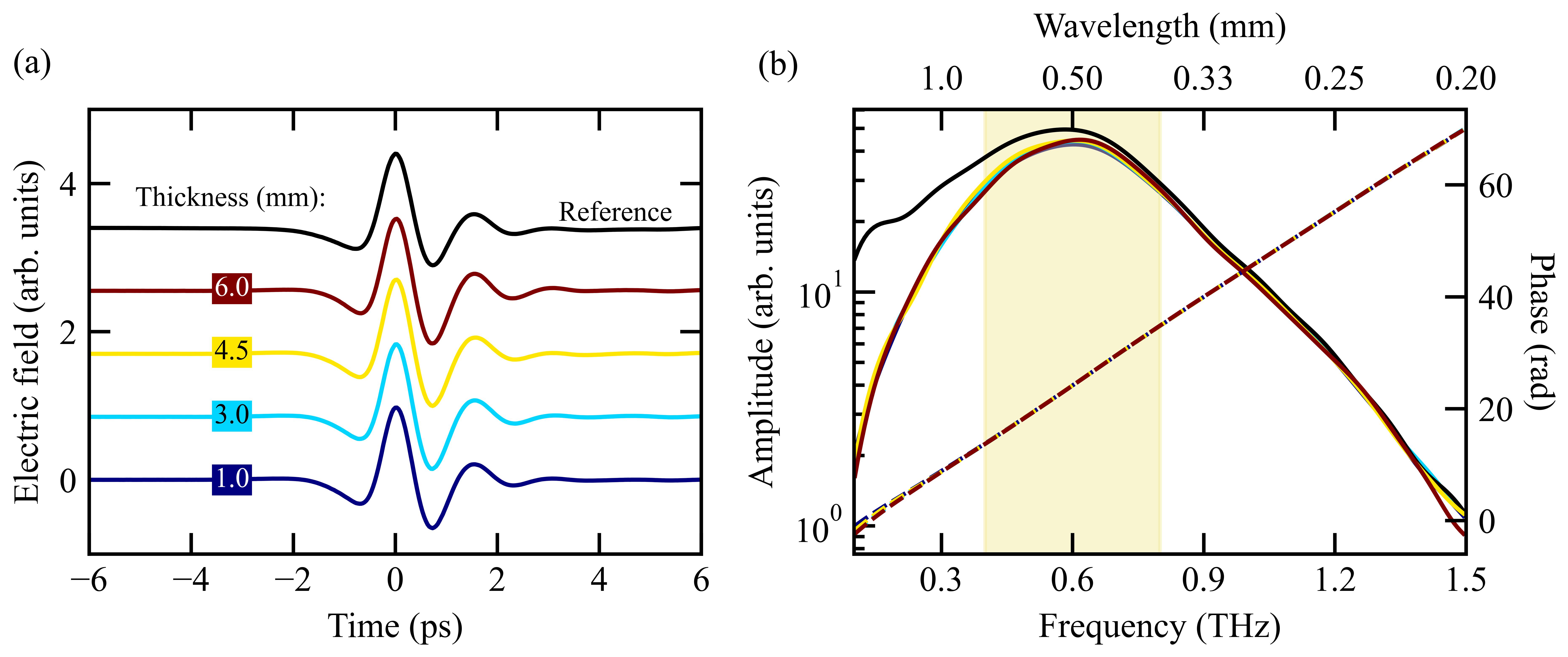}
\caption{Effect of aperture thickness on THz transmission for a fixed diameter of $D=\qty{3}{\mm}$. \textbf{(a)} Time-domain electric field profiles for four different thicknesses, compared with a reference pulse acquired without an aperture (black curve). All pulses exhibit similar amplitude and shape, with no evidence of delayed features or internal reflections. \textbf{(b)} Corresponding amplitude and phase spectra. The aperture spectra are nearly identical across all thicknesses, showing the expected low-frequency attenuation due to beam truncation. The phase curves remain well aligned with the reference, confirming the absence of cavity-induced effects or dispersive distortions.}\label{fig6}
\end{figure*}

These deviations arise from spatial filtering and spectral distortion caused by beam truncation. As the aperture narrows, long-wavelength components are increasingly attenuated, and phase shifts appear in the transmitted signal. The slab model interprets these effects as effective material dispersion and absorption, leading to an increase in $\kappa$ at low frequencies and a broadband upward shift in $n$.

Importantly, these effective parameters do not represent physical material properties but rather artifacts introduced by the aperture geometry. In practical THz-TDS measurements, samples placed on small apertures may yield retrieved optical constants contaminated by similar effects. This highlights the need for careful aperture selection and beam characterization when accurate low-frequency optical property extraction is critical.

\subsection{Influence of Aperture Thickness on THz Transmission}

We now assess whether aperture thickness influences the transmitted signal by analyzing a complementary dataset acquired with a fixed aperture diameter of $D=\qty{3}{\mm}$ and varying thicknesses: \qty{1}{\mm}, \qty{3}{\mm}, \qty{4.5}{\mm}, and \qty{6}{\mm}. Figure~\ref{fig6}a shows the time-domain electric field profiles for each case, along with the reference pulse acquired without an aperture (black line). All pulses exhibit nearly identical amplitude, arrival time, and overall shape, with no secondary features indicative of internal reflections or cavity-like interference.

The corresponding frequency-domain spectra are shown in Fig.~\ref{fig6}b. The amplitude and phase spectra for all aperture thicknesses are virtually indistinguishable, confirming that the transmitted signal remains unaffected across this range. As expected, all curves display the low-frequency attenuation characteristic of spatial filtering through a \qty{3}{\mm} aperture, while the reference spectrum (black) maintains higher intensity in this region. Importantly, the phase spectra for all aperture thicknesses remain well aligned with the reference, further ruling out any significant distortion or delay.

These results indicate that, for the geometry and frequency range considered here, aperture thickness does not introduce measurable effects in the THz response. This supports the assumption made in our transmission model, that thick apertures behave as uniform dielectric slabs without resonant or cavity-induced artifacts.

\subsection{Aperture-Dependent THz Response of a Quantum Material}

To assess the practical implications of the aperture-dependent effects described above, we performed THz-TDS measurements on a representative quantum material using sample holders with different aperture diameters: \qty{1}{\mm}, \qty{2}{\mm}, and \qty{3}{\mm}. This case study aims to validate how the geometric filtering observed in empty-aperture measurements manifests in a realistic sample spectrum, directly testing the impact of aperture size on spectral fidelity.

The sample is a natural bulk clinochlore crystal, shaped as a 140-µm-thick nearly square slab with lateral dimensions of approximately 4 mm. Clinochlore is a van der Waals mineral \cite{de_oliveira_high_2022} hosting a shear phonon mode near \qty{1.13}{\THz}, which we have recently shown to exhibit impurity-tuned optical anisotropy that reshapes the polarization state of transmitted THz waves~\cite{Kawahala2025} and displayed memory functionalities~\cite{doi:10.1021/acsami.4c19337}. Such properties make clinochlore a prototypical platform for developing two-dimensional devices with controllable phononic characteristics.

\begin{figure}[ht!]
\centering
\includegraphics[width=0.98\linewidth]{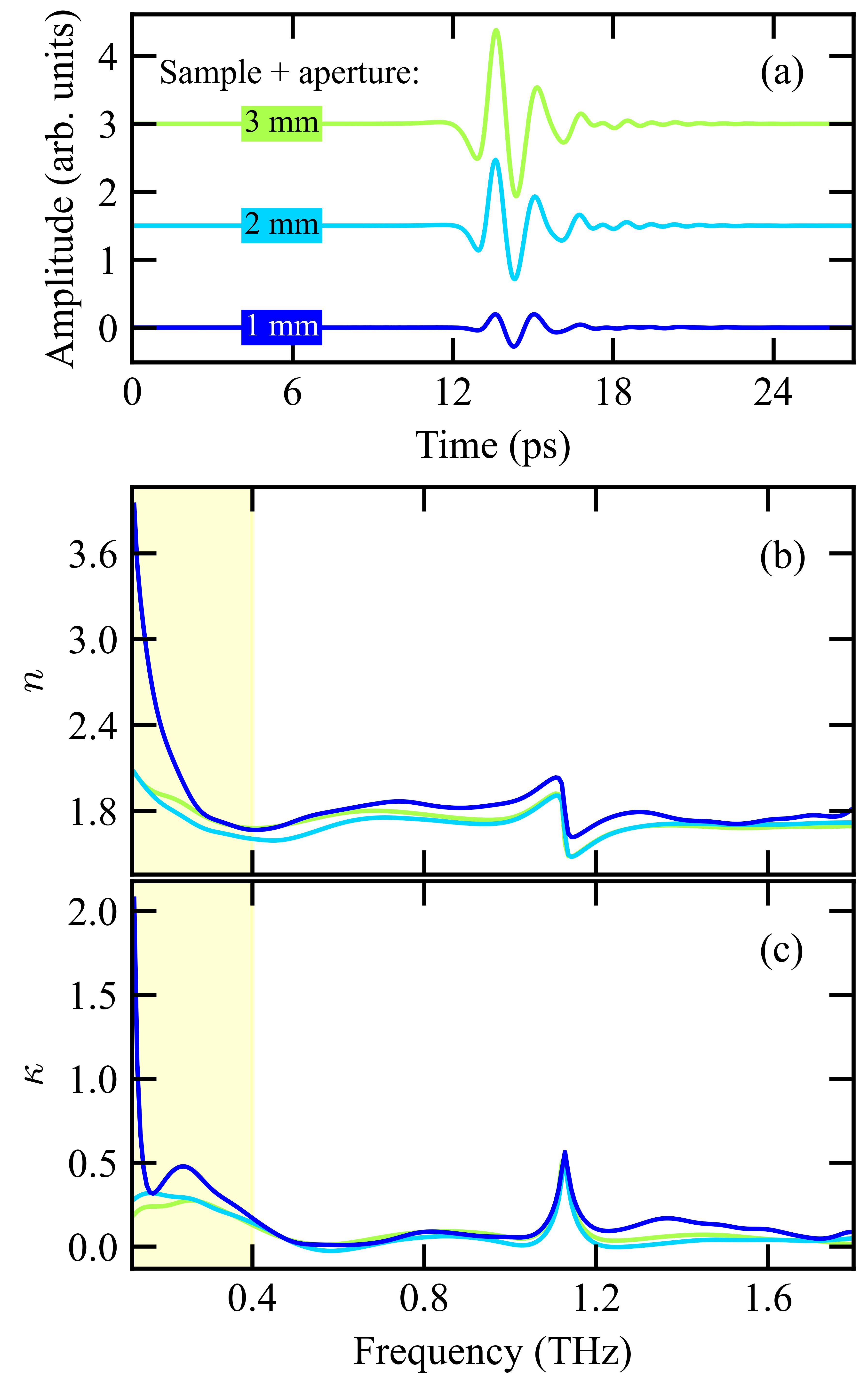}
\caption{THz-TDS measurements of the representative clinochlore sample performed with different aperture diameters. \textbf{(a)} Time-domain electric field waveforms transmitted through the sample using \qty{1}{\mm} (dark blue), \qty{2}{\mm} (light blue), and \qty{3}{\mm} (green) apertures. \textbf{(b)} Frequency-dependent real refractive index $n(\nu)$ and \textbf{(c)} extinction coefficient $\kappa(\nu)$, both retrieved by applying the transmission model described in the Methods section and using the corresponding empty-aperture measurement as reference.}\label{fig7}
\end{figure}

Figure~\ref{fig7} summarizes the measurements performed with the three aperture diameters. Panel (a) shows the time-domain electric field waveforms transmitted through the clinochlore sample, where the overall amplitude strongly depends on aperture size. The pulse measured with the \qty{2}{\mm} aperture exhibits a peak-to-peak amplitude approximately \qty{70}{\percent} that of the \qty{3}{\mm} case, while for the \qty{1}{\mm} aperture the signal drops to about \qty{20}{\percent}. Panels (b) and (c) present the frequency-dependent real refractive index $n(\nu)$ and extinction coefficient $\kappa(\nu)$, respectively, retrieved with the transmission model from the Methods section using, for each aperture diameter, the corresponding empty-aperture measurement as the reference. Although derived here to describe the apertures as air-filled slabs, the same formalism (Eq.~\ref{eq:2-8}) was originally developed for bulk samples~\cite{Neu2018}, such as the one analyzed in this section.

The most pronounced differences among the curves occur at low frequencies. For the \qty{1}{\mm} aperture, both $n$ and $\kappa$ deviate sharply below \qty{0.3}{\THz}, where the retrieved values increase unphysically due to the strong attenuation and phase uncertainty in this range. An additional artifact appears in $\kappa(\nu)$ between \qty{0.2}{\THz} and \qty{0.3}{\THz}, manifesting as a spurious peak absent from the larger-aperture spectra. This feature highlights how small apertures can distort the retrieved optical constants and potentially lead to misinterpretation of material excitations.

Above approximately \qty{0.4}{\THz}, the influence of aperture diameter becomes minor. The three spectra converge, with the background refractive index for the \qty{1}{\mm} case being roughly \qty{5}{\percent} higher than those for the larger apertures. The extinction coefficients remain nearly identical across all cases, except above \qty{1.2}{\THz}, where the \qty{1}{\mm} curve shows a modest increase. These results confirm that small-aperture effects primarily impact the low-frequency region of the spectrum, where accurate retrieval of optical parameters is most critical for characterizing low-energy excitations in quantum materials.

\section{Conclusion}\label{sec:conclusion}
We presented a systematic investigation of how circular apertures influence terahertz transmission in a standard THz-TDS setup. By analyzing both time-domain and frequency-domain data for apertures of varying diameters, we characterized the spatial and spectral filtering imposed by the aperture geometry and demonstrated a practical method for beam waist estimation based on transmitted intensity. The retrieved waist radius $w_0=\qty{2.36(21)}{\mm}$ ($\textrm{FWHM}=\qty{2.78(25)}{\mm}$) is consistent with typical focused THz beams in free-space systems~\cite{Reiten2003, Gaspar2023}, providing a quantitative benchmark for system alignment and spatial overlap.

Our results reveal that decreasing the aperture diameter progressively suppresses low-frequency components of the transmitted THz signal, while higher-frequency components are comparatively preserved. This frequency-selective attenuation reflects the narrowing of the beam waist with increasing frequency, which allows higher-frequency components to pass more efficiently through smaller openings. For very small apertures---particularly those approaching or below \qty{1}{\mm}---these effects become especially severe, accompanied by noticeable phase distortions that compromise the accuracy of frequency-domain analysis.

Such effects are especially relevant for the spectroscopy of quantum materials, whose dimensions are often limited and require tightly focused beams. Many of their key low-energy excitations---such as phonons, magnons, and free-carrier responses---occur at frequencies below \qty{1}{\THz}, where amplitude suppression and phase errors from small apertures can significantly degrade spectral fidelity. These distortions can obscure physical features and bias the retrieval of parameters like carrier mobility or scattering rate, particularly when modeling broadband responses such as Drude-like behavior near zero frequency.

We further applied an analytical transmission model to extract effective optical parameters---transmittance, refractive index, and extinction coefficient---from the measured spectra. The results showed increasing deviations from free-space values as the aperture diameter decreased, reinforcing the need for caution when interpreting transmission data obtained through narrow spatial filters. This was further confirmed by THz-TDS measurements of a representative quantum material, which revealed analogous aperture-dependent distortions in the retrieved optical parameters, validating the general conclusions of this study. Our findings establish a practical lower bound for aperture size in THz-TDS configurations based on focused beam waist, and suggest that accurate optical characterization requires careful control over aperture geometry.

Finally, we verified that aperture thickness, within the range tested (\qtyrange{1}{6}{\mm}), does not introduce measurable spectral or temporal distortions. This supports the common assumption that thick, non-resonant apertures act as simple dielectric slabs, validating their treatment in simplified transmission models.

In summary, this work offers practical guidance for optimizing aperture selection and sample mounting in THz spectroscopy, as well as a compact, non-invasive method for characterizing THz beam profiles—directly applicable to experimental setups employed in the study of low-dimensional quantum materials.

\backmatter


\bmhead{Acknowledgements}
The authors thank Dr. Ingrid D. Barcelos and Dr. Raphaela de Oliveira for supplying the clinochlore mineral. 

\bmhead{Author contributions}
N.M.K. and F.G.G.H. conceived the project. N.M.K. developed the THz-TDS setup used in the experiments. E.D.S. and N.M.K. performed the measurements, and L.O.D. and E.D.S. carried out the data analysis. L.O.D. and E.D.S. prepared the original draft of the manuscript, which was revised and edited by N.M.K. F.G.G.H. supervised the project and contributed to its overall direction. All authors discussed the results and contributed to the final version of the manuscript.

\bmhead{Funding}
This work was supported by the São Paulo Research Foundation (FAPESP), Grants Nos. 2021/12470-8 and 2023/04245-0. L.O.D. acknowledges support from the Programa Unificado de Bolsas de Estudo para Apoio à Formação de Estudantes of the University of São Paulo (PUB-USP). E.D.S. acknowledges financial support from Grant No. 407815/2022-8 of the National Council for Scientific and Technological Development (CNPq) and from FAPESP Grant No. 2025/02029-3. N.M.K. acknowledges support from FAPESP Grant No. 2023/11158-6. F.G.G.H. acknowledges support from CNPq Grant No. 306550/2023-7.

\bmhead{Data availability}
All data that support the conclusions of this work are contained within the article. Additional data will be made available upon request.

\bmhead{Corresponding author}
Correspondence to Felix G. G. Hernandez.

\section*{Declarations}

\textbf{Competing interests} The authors declare no competing interests.





\bibliography{sn-bibliography}

\end{document}